\documentclass[preprint,superscriptaddress,showpacs,preprintnumbers,amsmath,amssymb]{revtex4-1}

\usepackage{amsmath}
\usepackage{amssymb}
\usepackage{graphicx}
\usepackage{morefloats}
\usepackage[usenames,dvipsnames]{xcolor}
\usepackage{blkarray}
\usepackage{verbatim}
\usepackage{hyperref}
\usepackage{color}
\usepackage{enumerate}
\usepackage{multirow}
\usepackage{twoopt}
\usepackage{array}
\usepackage{alltt}

\begin{document}

\title{Solving Optimization Problems by the Public Goods Game}

\author{Marco Alberto Javarone}

\email{marcojavarone@gmail.com}

\address{
School of Computer Science, University of Hertfordshire, Hatfield AL10 9AB, UK}
\address{Dept. of Mathematics and Computer Science, University of Cagliari, 09123 Cagliari Italy}

\date{\today}

\begin{abstract}
We introduce a method based on the Public Goods Game for solving optimization tasks. 
In particular, we focus on the Traveling Salesman Problem, i.e. a NP-hard problem whose search space exponentially grows increasing the number of cities.
The proposed method considers a population whose agents are provided with a random solution to the given problem. In doing so, agents interact by playing the Public Goods Game using the fitness of their solution as currency of the game. Notably, agents with better solutions provide higher contributions, while those with lower ones tend to imitate the solution of richer agents for increasing their fitness.
Numerical simulations show that the proposed method allows to compute exact solutions, and suboptimal ones, in the considered search spaces.
As result, beyond to propose a new heuristic for combinatorial optimization problems, our work aims to highlight the potentiality of evolutionary game theory beyond its current horizons.
\end{abstract}

\maketitle
Nowadays, evolutionary game theory~\cite{nowak01, perc01,moreno01,moreno02,tomassini01,moreno03} represents a field of growing interest in different scientific communities, as biology~\cite{traulsen01,nowak02} and social science~\cite{perc02}.
Notably, identifying strategies and methods for triggering cooperative behaviors~\cite{nowak03}, modeling biological phenomena~\cite{nowak04} and studying the effects of social influences~\cite{perc02,masuda01,javarone01} constitute some of the major aims in this field.
On the other hand, the Darwinian concept of evolution, underlying the dynamics of evolutionary games, represents a powerful and inspiring source also in the field of natural computing~\cite{natural01}.
In the last years, many evolutionary algorithms~\cite{holland01,goldberg01} have been proposed for solving optimization problems~\cite{krentel01,dorigo03,kellerer01}, as for instance genetic algorithms~\cite{holland01} and ant colony heuristics~\cite{dorigo01}.
Remarkably, optimization problems have been widely investigated also within the realm of statistical physics~\cite{zecchina01,zecchina02,krzakala01,zamponi01,amit01,hopfield01,kirkpatrick01,mezard01}, where theoretical physics and information theory meet forming a powerful framework for studying complex systems~\cite{anderson01,miguel01}.
For instance, a statistical physics mindset in combinatorial optimization problems emerges when the set of feasible solutions, of problems like the Traveling Salesman Problem~\cite{dorigo02,agliari02} (TSP hereinafter), is represented in terms of an energy landscape. In doing so, the searching of a solution corresponds to the searching of a minimum of free energy, in a landscape whose global minimum, i.e. the deepest valley, corresponds to the optimal solution of the problem.
Different models as the Curie-Weiss~\cite{wolski01} and spin glasses~\cite{parisi02,zamponi02} have an energy that can be studied by the Landau formulation of phase transitions~\cite{huang01}. These models are successfully adopted for facing different issues as opinion dynamics~\cite{galam01}, information retrieval~\cite{amit01,barra01,barra02}, optimization tasks~\cite{zecchina01,javarone02} and learning processes~\cite{amit01}.
Using the metaphor of the energy, heuristics like genetic algorithms~\cite{holland01} and swarm logics~\cite{dorigo01}, implement strategies as genetic recombination, mutation, and collective motion, for surfing the energy landscape with the aim to reach one of the more deep valleys in few time, i.e. one of the suboptimal solutions of a problem. Therefore, parameters as the mutation rate, used in genetic algorithms, can be compared to physical parameters, e.g. the temperature of a system.
In this work, we adopt a mechanism based on the partial imitation~\cite{javarone02}: when an agent interacts with another one having a higher fitness, the former imitates a part of the latter's solution. For example, in the TSP, the weaker agent imitates only a part of the path traveled by a stronger opponent. In doing so, agents are able to generate solutions over time, with the aim to achieve the optimal one.
In physical terms, a partial imitation can be interpreted as a slow cooling process of a spin particle system, where the slowness comes from an imitative dynamics that is only `partial' (i.e. only few entries of a solution array are imitated).
Our model considers an agent population, whose interactions are based on the Public Goods Game (PGG hereinafter). As we know from evolutionary game theory (EGT hereinafter)~\cite{perc03}, the outcomes of the classical PGG are affected by a parameter defined synergy factor $r$, used for supporting cooperators. Here, as shown below, this parameter (i.e. $r$) has a marginal role, however what is relevant for our investigations is that an ordered phase (i.e. the prevalence of a species in the population) can be reached by an opportune tuning of its value. Usually, in EGT models, a species indicates a set of agents with the same strategy, e.g. cooperation, whereas in the proposed model a species corresponds to a set of agents having the same solution of a problem.
In general, ordered phases entail all agents have the same state (or strategy in EGT), i.e. in physical terms all spins are aligned in the same direction. Here, the magnetization is a useful parameter that allows to measure the state of order of a system and, in the ordered case, it has a value equal to $\pm 1$.
Dealing with neural networks, and in general with spin glasses, it is possible to introduce a gauge for the magnetization so that its value goes to $\pm 1$ when the spin alignment (i.e. agent states) follows a particular pattern. For instance, in the case of the TSP, a pattern can be a specific sequence of cities. The mentioned gauge is defined Mattis magnetization~\cite{amit01}, and it reads $M_m = \frac{1}{n} \sum_i \epsilon_i s_i$ with $\epsilon_i$ value in the $i$-th position of the pattern, $s_i$ value of the spin in the same position of a signal $S$ of length $n$. As we can observe, when spins are perfectly aligned with a pattern $\epsilon$, the Mattis magnetization is equal to $1$.
In the proposed model, we introduce a similar approach. In particular, each agent is provided with a random solution of the TSP (i.e. an array of cities representing a possible solution), and the order is reached when all agents hold the same solution. Therefore, in our case, the value of $M_m$ is computed assigning the value of $+1$ when a city has the same position both in the pattern of reference (i.e. the known optimal solution of a TSP problem), and in the solution array computed by an agent, otherwise the value is $-1$. It is worth to recall that the utilization of the Mattis magnetization, as measure for the performance of our model, can be adopted only when the optimal solution is known in advance.
Since our agents interact by the PGG, the modification of their solution occurs during the phase of the game usually defined as `strategy revision phase'~\cite{tomassini01}, that in our case is renamed as `solution revision phase'.
Furthermore, our agents use their fitness as currency of the game, so that their payoff depends on the quality of their solution and on those of their opponents. 
We performed several numerical simulations to evaluate the quality of our method considering the TSP as reference, i.e. a famous NP-hard problem.
Results show that the PGG can be successfully adopted for developing new heuristics, opening the way to investigations that cross the current fences of EGT.
\section{Model}\label{sec:model}
Before introducing the proposed model, let us recall the basic dynamics of the PGG. The latter considers a population with $N$ agents and two possible strategies: cooperation and defection. Cooperators contribute to a common pool with a coin, while defectors contribute nothing or, as in our case, provide a partial contribution (i.e. a coin whose value is lower than that of coins provided by cooperators). Then, the total amount of coins is enhanced by a synergy factor $r$ (whose value is greater than $1$), and the resulting value is equally divided among all agents (no matter their strategy).
In doing so, each agent receives a payoff which reads
\begin{equation}\label{eq:pgg_payoff}
\begin{cases}
\pi^{c} = r \frac{\sum_{i=1}^{N^c} c_i}{G} - c\\
\pi^{d} = r \frac{\sum_{i=1}^{N^c} c_i}{G}
\end{cases}
\end{equation}
\noindent with $N^c$ number of cooperators, $G$ amount of agents involved in the game (i.e. size of groups considered at each iteration that, usually, is much smaller than $N$), $c_i$ unitary contribution (we can set, without loss of generality, equal for all agents, i.e. $c_i = c = 1$), and $\pi^{c}$ and $\pi^{d}$ payoff of cooperators and defectors, respectively.
As the quantitative definition of the payoff suggests, defection is more convenient than cooperation, and it also represents the Nash equilibrium of this game.
The role of the synergy factor $r$ is promoting cooperation and, as demonstrated in previous investigations, its value may strongly affect the evolution of a population~\cite{perc03}. Remarkably, in square lattices, values of $r$ smaller than $3.75$ entail all agents become defectors, whereas higher values allow cooperators to survive and even to succeed (for $r \ge 5.49$).
As previously mentioned, the evolution of a population results from the process defined `strategy revision phase'. Notably, after each iteration, an agent has the opportunity to change its strategy by imitating that of a richer opponent. Here, the richness is related to the gained payoff.
In the proposed model we consider a well-mixed population, so that agents may freely interact with their opponents. Moreover, agents are provided with a random solution of a TSP (i.e. an array of cities). Notably, each solution is evaluated by a fitness $\eta$ computed as follows
\begin{equation}\label{eq:fitness}
\eta = \frac{Z - 1}{D}
\end{equation}
\noindent with $Z$ number of cities, and $D$ total distance of a path. In doing so, the fitness has a range $\eta \in [0,1]$. 
At each time step, one agent is randomly selected (say the $x$th) and plays the PGG with $4$ (randomly chosen) opponents, forming a group with $G = 5$ agents. Now, every agent of the group contributes with its fitness; then, as in the PGG, the total summation of contributions is enhanced by a synergy factor $r$ and, finally, equally distributed among all agents of the group.
It is worth noting that, in the proposed model, all agents always contribute. However, some agents provide a contribution higher/smaller than that of others. Therefore, 'below average contributors' (i.e. those having a low quality solution) can be considered as defectors~\cite{fehr01}.
According to this setting, the payoff reduces to one equation
\begin{equation}\label{eq:payoff_mod}
\pi_x = r \frac{\sum_{i=1}^{5} \eta_i}{G} - \eta_x
\end{equation}
\noindent with $\pi_x$ indicating the payoff of the $x$th agent, and $\eta_x$ its fitness (i.e. that corresponding to its solution).
Finally, the `strategy revision phase', in our model, is renamed `solution revision phase': the randomly selected agent computes the probability $\Pi_s$ to modify each entry of its solution by imitating that of its best opponent (if exists)
\begin{equation}\label{eq:adaptation}
\Pi_s = \frac{1}{1 + e^{\frac{\eta_x - \pi_x}{K}}}
\end{equation}
\noindent As in the PGG, $K$ represents the uncertainty in imitating an opponent (i.e. plays the role of temperature). Hence, setting $K = 0.5$ we implement a rational approach during the revision phase~\cite{perc03}. Therefore, the $x$th agent imitates with probability $\Pi_s$ each entry of the solution of its best opponent, if the latter has a greater or, at least, an equal fitness (otherwise the $x$th agent does not revise its solution).
Summarizing, given a TSP, we define a population whose agents at the beginning receive a random solution of the problem. Then, local interactions, based on the PGG, allow the population to converge towards a shared solution. From a local point of view, at each time step, a randomly selected agent (say $x$) plays the PGG with $4$ (randomly chosen) opponents, and computes its payoff (i.e. by Eq.(~\ref{eq:payoff_mod})). So, according to its fitness $\eta_x$ and to the gained payoff $\pi_x$, the $x$th agent computes the probability $\Pi_s$ to imitate the solution of its best opponent (say $y$, if exists). In particular, if $\eta_y \ge \eta_x$, the $x$th agent revises its solution, i.e. it imitates each entry of the solution of the $y$th agent with probability $\Pi_s$ (i.e. each entry is modified according to $\Pi_s$). The whole process is repeated until the population reaches an ordered phase (i.e. all agent share the same solution), or up to a limited number of time steps elapsed.
It is worth observing that as $\Pi_s$ goes to $1$, the imitation process entails one agent tends to copy the whole solution of its best opponent. In addition, we remark that when an agent performs a 'partial imitation', for instance to modify one city along its path, the same city cannot be visited twice (i.e. it can be present in only one cell of the solution array).
In order to clarify this point we provide a simple example. Let us consider an agent having the following solution: $[$ Paris, New York, London, Miami, Rome, Madrid $]$, that has to put in the third cell (now containing London) the city of Rome. Since currently Rome is in the fifth cell, the algorithm swaps the values for the third and fifth cells so that, after the whole process, the resulting array is: $[$ Paris, New York, Rome, Miami, London, Madrid $]$. Thus, repetitions are completely avoided, and all solutions generated according to the proposed heuristic are suitable solutions.
Finally, we deem relevant to emphasize the main differences between the PGG and the proposed model. First, in our model, the contributions provided by the agents correspond to their fitness, while in the PGG contributions just represent forms of cooperation to a common wealth. Second, the 'strategy revision phase' of the PGG, here renamed 'solution revision phase', entails an imitation process between two agents that can be complete or only partial. Moreover, the imitation probability (i.e.~\ref{eq:adaptation}) takes as input the payoff and the fitness of the same agent, i.e. the one that undergoes the revision process. Finally, a further important difference, between the PGG and the proposed model, is given by the number of possible ordered equilibria. Notably, in the PGG, the possible ordered phases can correspond to full cooperation, or full defection. Instead, in the proposed model, each suitable solution of a combinatorial optimization problem can be an ordered equilibrium that the population can reach.
\section{Results}\label{sec:results}
\begin{figure}[h]
\centering
\includegraphics[width=0.475\textwidth]{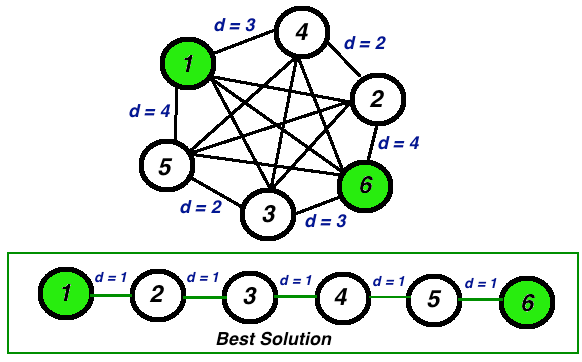}
\caption{\small General setting of the TSP considering $Z = 6$ cities forming a complete graph. Each node represents a city, and some distances are reported in blue, close to the related link. Then, the best solution is shown. Green nodes represent the starting and the landing ones.\label{fig:figure_0}}
\end{figure}

\begin{figure}[!ht]
\centering
\includegraphics[width=0.475\textwidth]{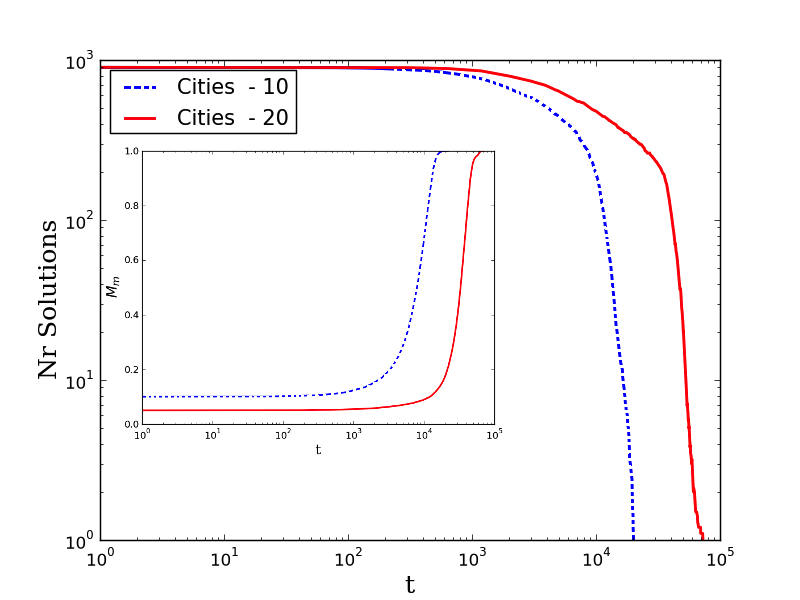}
\caption{\small Number of solutions over time in a population of $N = 900$ agents while solving a TSP with $10$ cities (blue dotted line) and $20$ cities (red line). The inset shows the related Mattis magnetization for the two cases (both successful). Results have been averaged over $100$ different attempts.\label{fig:figure_1}}
\end{figure}

\begin{figure}[!h]
\centering
\includegraphics[width=0.48\textwidth]{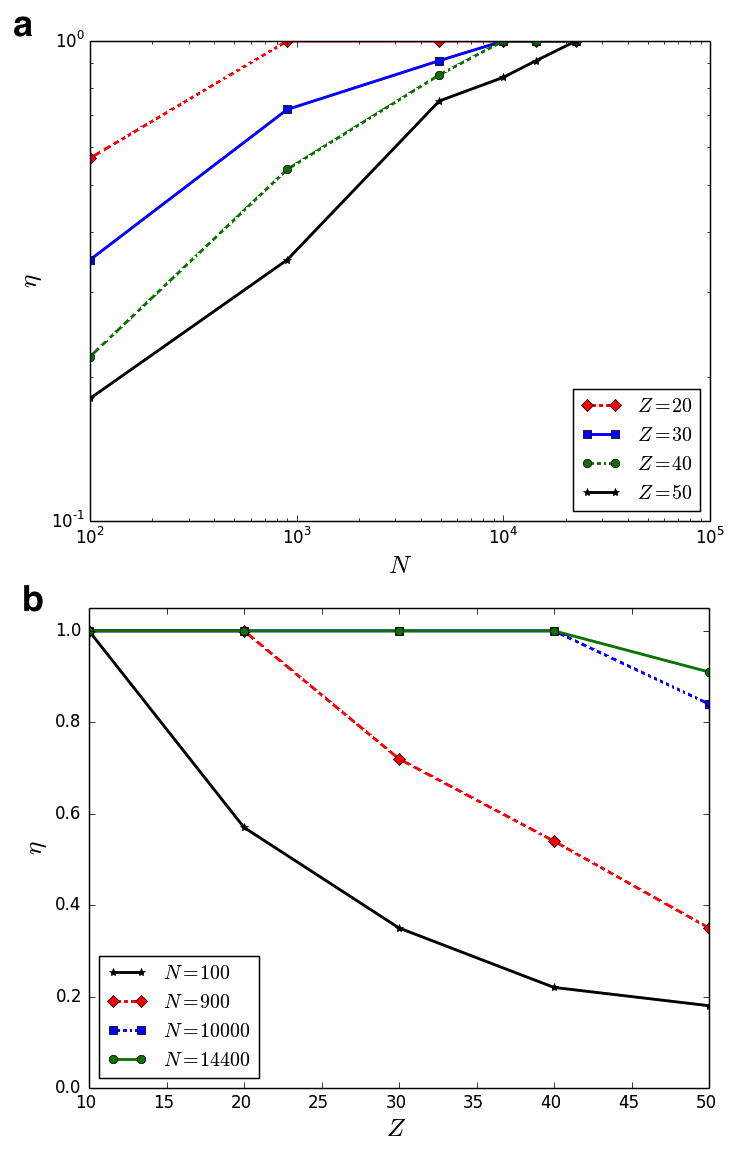}
\caption{\small \textbf{a)} Average fitness of the final solution in function of $N$ (i.e. the number of agents), for different values of $Z$ (i.e. the number of cities). \textbf{b)} Average fitness of the final solution on varying the number of cities, for different agents $N$. Results have been averaged over $100$ different attempts. \label{fig:figure_2}}
\end{figure}

\begin{figure}[h]
\centering
\includegraphics[width=0.48\textwidth]{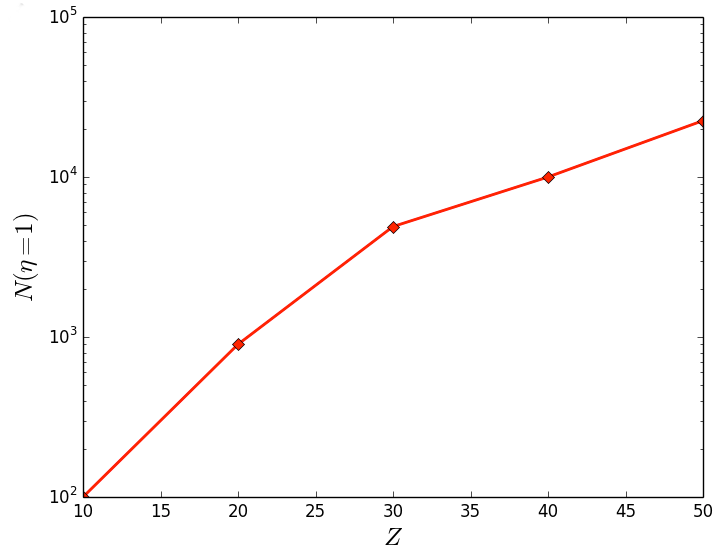}
\caption{\small Minimum number of agents to compute the optimal solution of a TSP on varying the number of cities $Z$. Results have been averaged over $100$ different attempts. \label{fig:figure_3}}
\end{figure}

\begin{figure}[h]
\centering
\includegraphics[width=0.48\textwidth]{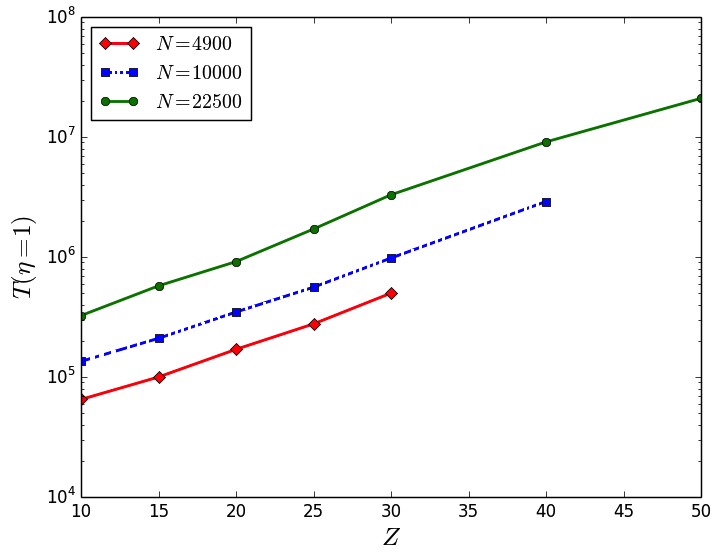}
\caption{\small Number time steps required for converging to the final (optimal) state on varying $Z$, for different population sizes $N$. Results have been averaged over $100$ different attempts. \label{fig:figure_4}}
\end{figure}

Numerical simulations have been performed considering a number of cities up to $Z = 50$ for defining the TSP.
Agents know the starting city and the landing one so, since each city can be visited only once, the number of feasible solutions is $(Z-2)!$. Moreover, without loss of generality, we consider that the distance between two close cities is always equal to one ---see Fig.~\ref{fig:figure_0}.
Eventually, we set the synergy factor to $r = 2$. We remind that in the present work we are not interested in studying phenomena as the evolution of cooperation, but we aim to evaluate if agents are able to converge towards an ordered phase, characterized by the existence of only one shared solution of a problem. Thus, the choice of setting $r = 2$ reflects this requirement, i.e. to use a value that in the PGG leads to an ordered phase (i.e. full defection in the specific case).
As illustrated in Fig.~\ref{fig:figure_1}, the ergodicity of the process always allows agents to converge to one common solution. Moreover, we are able to verify the quality of solutions both considering the related fitness and the Mattis magnetization (see the inset of Fig.~\ref{fig:figure_1}). In particular, the latter can be used when the solution of a problem is known in advance (as in our case).
An important relation to be considered is the one defined between the final average fitness and the size of the population $N$, studied on varying the amount of cities $Z$ ---see plot \textbf{a} of Fig.~\ref{fig:figure_2}. 
Moreover, as shown in plot \textbf{b} of Fig.~\ref{fig:figure_2}, it is worth noting that also good suboptimal solutions may be computed using a number of agents $N$ smaller than that required to compute the optimal one.
As expected, increasing $Z$ the average value of $\eta$ reduces (keeping fixed the number of agents $N$). On the other hand, as shown in Fig.~\ref{fig:figure_3}, it is worth highlighting that it is possible to find an opportune $N$ for each considered $Z$ in order to achieve the highest fitness (i.e. $\eta = 1$). We deem relevant to note that the number of agents to compute the best solution, i.e. $N(\eta = 1)$, is much smaller than the number of feasible solutions for each problem, therefore our method can be considered a viable heuristic for facing combinatorial optimization problems.
Eventually, we focused on the number of time steps to let the population converge, considering in particular the successful cases, i.e. those leading to the optimal solution ---see Fig.~\ref{fig:figure_4}. As expected, wide search spaces (e.g. $Z = 50$) require more time steps to let the population converge to the same final (and optimal) solution. Furthermore, increasing $N$ and keeping fixed $Z$, the number of time steps $T$ increases accordingly. 
These results are in full agreement with converging processes that can be observed in generic agent-based models, e.g. increasing the size of a population the number of time steps, required to let agents converge towards the same state, increases~\cite{redner01}.

\section{Discussion and Conclusion}\label{sec:conclusion}
In this work we show that evolutionary games as the PGG can be, in principle, applied also for solving combinatorial optimization problems.
In particular, the order-disorder phase transition occurring in a population interacting by the PGG can be adopted for letting the population converge towards a common solution of a problem. Notably, a solution plays the same role of a strategy in the classical PGG, and the order is reached by implementing a mechanism of `partial imitation'~\cite{javarone02}. 
The latter allows agents with a weak solution to partially imitate stronger (i.e. richer) opponents. From a physical perspective, this mechanism can be viewed as a slow cooling process that triggers the emergence of solutions over time, whereas the ergodicity of the process allows the population to reach an absorbing state of full order. In doing so, an ordered phase entails all agents share the same solution.
Under the hypothesis that an evolutionary dynamics driven by the payoff, i.e. rational, may constitute the base for solving difficult problems as the TSP, we performed several numerical simulations by considering a well-mixed population.
Although we implemented a simplified version of the TSP, with a limited number of cities, it is worth highlighting that results indicate that the proposed model allows to compute the optimal solution in all considered search spaces. Moreover, even using a reduced number of agents, it is possible to compute a good suboptimal solution.
Furthermore, we note that even introducing spatial constraints in the TSP (e.g. a missing link between two cities), the algorithm is able to face the problem, once the driveability of the graph is known (as shown in Fig.~\ref{fig:figure_0}).
Therefore, in the light of the achieved outcomes, we deem relevant to further investigate the potential of evolutionary games in optimization problems, then enlarging the domain of applications of EGT.
However, it is important to emphasize that in order to really appreciate the quality of the proposed model as algorithm for solving the TSP, further investigations are required. In particular, those for comparing the performances with other heuristics, as genetic algorithms (see Appendix I). On the other hand, we remark that our results indicate a clear relation between the size of a population and the complexity of the faced problem. This last observation constitutes a first, even if theoretical, advantage of our method respect to the others because, as far as we know, similar relations are not available for other methods.
Now, from the point of view of EGT, there are two important observations. First, the synergy factor has a marginal role in the proposed model. We recall that, for the aims of our work, we are interested in allowing the population to converge towards an ordered state. On studying the PGG, the synergy factor is fundamental because, as before mentioned, some values may lead a population towards a steady-state of coexistence between cooperators and defectors. Therefore, since here we have to avoid similar scenarios, in principle, every value of the synergy factor that supports a generic state of full order can be adopted. 
At the same time, we think that the synergy factor should not be too high, otherwise it might generate problems when computing transition probabilities during the 'solution revision phase'. 
In particular, as indicated in Eq.(~\ref{eq:adaptation}), the fitness and the payoff are compared when evaluating whether one agent has to change its strategy. Thus, we suggest to use small values, like the one we adopted (i.e. $r = 2$). The second observation is related to the identification of defectors. Notably, here we refer to the PGG, i.e. a simple game with two strategies: cooperation and defection. In the classical version, cooperators contribute with a coin, while defectors do not contribute. However, as reported in~\cite{fehr01}, when the amount of contributions is not set to a specific value (e.g. a coin of unitary value), those agents that contribute with a below-average contribution can be considered as defectors. 
To conclude, the proposed heuristic shows that cooperative dynamics, leading from disordered to ordered states, may constitute the basic mechanism for implementing optimization algorithms.
\section*{Acknowledgments}
The author wishes to thank Adriano Barra for his helpful comments, and Matteo Matteucci for his useful suggestions. In addition, he would like to acknowledge support by the H2020-645141
WiMUST project and by National Group of Mathematical Physics (GNFM-INdAM).

\section*{Author contribution statement}
The author defined and developed the present work, and performed all related analyses.

\section*{Appendix I}
Here, we report results of a comparative analysis between the proposed method and two other heuristics: genetic algorithms~\cite{goldberg01} and social imitation~\cite{javarone02}. The former constitutes one of the most famous methods adopted in combinatorial optimization problems, while the second allows to evaluate the differences between the proposed strategy and one based on a simple imitative mechanism (based on the fitness).
Before to show a comparative table, we briefly summarize how the social imitation strategy works, and how the genetic algorithm has been implemented.
Let us start with the social imitation algorithm, implemented according to the following steps:
\begin{enumerate}
\item Define a population with $N$ agents, and assign each one a random solution for the considered TSP;
\item Compute the fitness $\eta$ of each agent (i.e. the goodness of its solution);
\item Compute the number of different solutions (say $K$) in the population;
\item IF $K > 1$:\\ \_\_$(i)$ randomly select two agents ($x$ and $y$) having different solutions:\\
\_\_\_\_ IF $\eta_x \le \eta_y$ perform a Solution Revision Phase (see below);\\ \_\_\_\_ ELSE REPEAT from $(i)$. \\ \_\_$(ii)$ REPEAT from $(3)$;\\\\
ELSE STOP.
\end{enumerate}
The Solution Revision Phase is composed of the following steps
\begin{enumerate}[a]
\item Randomly select a position, say $z$, (i.e. an entry in the solution array) of the $x$-th agent's solution;
\item Check that the value in $z$ be different between the two selected agents, otherwise repeat from $(a)$;
\item Compute the position, say $w$, in the $x$-th agent's solution containing the value in position $z$ of the $y$-th agent's solution;
\item Exchange in the $x$-th agent's solution the values contained in positions $z$ and $w$.
\end{enumerate}

Genetic Algorithms can be implemented in several ways. In this work, we consider a simple definition:
\begin{enumerate}
\item Define a population with $N$ genes, assign each one a random solution for the considered TSP, and define a maximum number of iterations $I$;
\item While the best fitness in the population is smaller than $1$, or the number of iterations is smaller than $I$:
\item \_ Compute the fitness $\eta$ of each gene (i.e. the goodness of its solution);
\item \_ Select the best half of the population according to fitness;
\item \_ Generate two new solutions for each couple of genes, defined among the set computed at the previous step;
\item \_ Apply the random mutation, to each gene, with probability $p_m$;
\end{enumerate}
We set to $0.1$ the probability $p_m$ (i.e. the random mutation), and to $30k$ the maximum number of iterations $I$. In addition, we emphasize that the crossover operator has been defined by cutting each gene parent (i.e. solution) in two different points, so generating an offspring by using the central part of one parent and the side parts of the other parent. In the case this process generates not viable solutions (e.g. in the presence of repetitions), the duplicates are removed for adding the missing cities.
Table~\ref{tab:table_a} shows the number of agents (or genes for the genetic algorithm) for computing the optimal solution on varying the number of cities, the average number of time steps required to complete a simulation and, when smaller than $1$, the average fitness.
\begin{table*}[h]
\centering
\caption{Performance comparison on varying the number of cities (Z): proposed method (PGG), Social Imitation (SI), and Genetic Algorithm (GA). $N$ indicates the minimum number of agents (genes for GA) used to solve the problem, and $<T>$ indicates the average number of time steps required. The average fitness $<\eta>$ is indicated only when smaller than $1$, although the best value computed considering all attempts is $1$ (i.e. the optimal solution has not been always computed). }\label{tab:table_a}
    \begin{tabular}{ | p{4cm} | p{4cm} | p{4cm} | p{4cm} |}
    \hline
    Z & PGG & SI & GA \\ \hline
    10 & $N = 100$  $ | $  $<T>= 1K$ & $N = 60$ $ | $ $<T>= 8K$& $N = 100$ $ | $ $<T>= 27$ \\ \hline
    15 & $N = 900$ $ | $   $<T>= 17K$ & $N = 100$ $ | $ $<T>= 40K$& $N = 100$ $ | $ $<T>= 200$ \\ \hline
    20 & $N = 900$ $ | $   $<T>= 29K$ & $N = 270$ $ | $ $<T>= 500K$& $N = 100$ $ | $ $<T>= 1.3k$ \\ \hline
    25 & $N = 4900$ $ | $ $<T>= 277K$ & $N = 500$ $ | $ $<T>= 1.5M$& $N = 100$ $ | $ $<T>= 4.1k$ \\ \hline
    30 & $N = 4900$ $ | $ $<T>= 500K$ & $N = 700$ $ | $ $<T>= 5.5M$& $N = 100$ $ | $ $<T>= 13.2k$ \\ \hline
    35 & $N = 4900$ $ | $ $<T>= 820K$ & $N = 1000$ $ | $ $<T>= 15.5M$& $N = 100$ $ | $ $<T>= 17.5k$ \\ \hline
    40 & $N = 10000$ $ | $ $<T>= 3M$ & $N = 1200$ $ | $ $<T>= 40M$&  $N = 200$ $ | $ $<T>= 23k$ $ | $ $<\eta>= 0.76$ \\ \hline
    50 & $N = 22500$ $ | $ $<T>= 21M$ & $N = 1600$ $ | $ $<T>= 360M$& $N = 200$ $ | $ $<T>= 28.5k$ $ | $ $<\eta>= 0.61$\\ \hline
    
    \hline
    \end{tabular}
   
\end{table*}

According to these results, we observe that the proposed method requires the highest number of agents to solve a TSP. However, if compared to the SI algorithm, our approach is much more faster (see the average number of time steps $<T>$) than SI. Therefore, this result seems to suggest that combining the 'game mechanism' in an imitation process makes sense for solving optimization problems.
The genetic algorithm is the one that required the smallest number of agents, and the smallest amount of time to complete a simulation. At the same time, it is important to observe that the genetic algorithm has a synchronous dynamics (while our method and SI are asynchronous), i.e. during the same time step, all agents are involved for generating offsprings and updating their solution (according to the random mutation mechanism). Therefore, further analyses are required for a complete time comparison. However, it seems that the genetic algorithm is the fastest one. Nevertheless we found that, considering $20$ different simulation runs, the average fitness of the best solution (found in the gene population) is smaller than $1$ when $Z \ge 40$. Hence, the genetic algorithm must be run several time for each task, saving the best solution.
To conclude, according to this analysis, we report that a genetic algorithm constitutes the best choice for solving simple problems (i.e. with few cities), or for computing in few time a good suboptimal solution with many cities. On the other hand, when the number of cities increases, the proposed method allows to reach a higher fitness in a number of attempts smaller than that required by a genetic algorithm, and to compute the optimal solution within a number of time steps much more smaller than that required by the social imitation strategy.

\end{document}